# Title:

**Uncovering low-dimensional, miR-based signatures of acute myeloid and lymphoblastic leukemias with a machine-learning-driven network approach**

## Authors and Affiliations:

Julián Candia[1,2,3], Srujana Cherukuri[3,4], Yin Guo[3], Kshama A. Doshi[3], Jayanth R. Banavar[2], Curt I. Civin[3], Wolfgang Losert[2]

[1] Center for Human Immunology, Autoimmunity and Inflammation, National Institutes of Health, Bethesda, MD 20892, USA
[2] Department of Physics, University of Maryland, College Park, MD 20742, USA
[3] Center for Stem Cell Biology & Regenerative Medicine, Departments of Pediatrics and Physiology, University of Maryland School of Medicine, Baltimore MD 21201, USA
[4] Noble Life Sciences, 22 Firstfield Rd, Gaithersburg, MD 20878, USA

## Corresponding Author:

Julián Candia
Address: Center for Human Immunology, Autoimmunity and Inflammation, National Institutes of Health, 10 Center Drive, 10/7N115, Bethesda, MD 20892, USA
Email: julian.candia@nih.gov

## Abstract

Complex phenotypic differences among different acute leukemias cannot be fully captured by analyzing the expression levels of one single molecule, such as a miR, at a time, but requires systematic analysis of large sets of miRs. While a popular approach for analysis of such datasets is principal component analysis (PCA), this method is not designed to optimally discriminate different phenotypes. Moreover, PCA and other low-dimensional representation methods yield linear or non-linear combinations of all measured miRs. Global human miR expression was measured in AML, B-ALL, and T-ALL cell lines and patient RNA samples. By systematically applying support vector machines to all measured miRs taken in dyad and triad groups, we built miR networks using cell line data and validated our findings with primary patient samples. All the coordinately transcribed members of the miR-23a cluster (which includes also miR-24 and miR-27a), known to function as tumor suppressors of acute leukemias, appeared in the AML, B-ALL and T-ALL centric networks. Subsequent qRT-PCR analysis showed that the most connected miR in the B-ALL-centric network, miR-708, is highly and specifically expressed in B-ALLs, suggesting that miR-708 might serve as a biomarker for B-ALL. This approach is systematic, quantitative, scalable, and unbiased. Rather than a single signature, our approach yields a network of signatures reflecting the redundant nature of biological signaling pathways. The network representation allows for visual analysis of all signatures by an expert and for future integration of additional information. Furthermore, each signature involves only small sets of miRs, such as dyads and triads, which are well suited for in depth validation through laboratory experiments. In particular, loss- and gain-of-function assays designed to drive changes in leukemia cell survival, proliferation and differentiation will benefit from the identification of multi-miR signatures that characterize leukemia subtypes and their normal counterpart cells of origin.

# 1. Introduction

Interdisciplinary research at the crossroads of physics, biology, mathematics, and informatics are best exemplified by complex network science [1-3]. From early efforts to describe the complex, large-scale, networked structure of metabolic, transcription regulatory and protein-protein interaction networks [4-8], the emergent network biology field has evolved to play a key role in cancer systems biology and systems biology at large [9-11]. Within this broad context, here we develop a framework that naturally integrates machine learning and complex network ideas into microRNA cancer biology.

MicroRNAs (miRs) are a class of short noncoding RNAs that target messenger RNAs to regulate gene expression post-transcriptionally. Expression of miRs is altered in acute leukemias [12], and miR signatures for different leukemia types have been found either by using large-scale clustering approaches [13,14], which aim at identifying large, highly correlated groups of differentially expressed miRs and thus inferring relevant pathways, or by focusing on single miRs to find statistically significant population differences [12,15-17]. **This study addresses the gap between the encompassing view of the entire miR landscape offered by clustering techniques, and the narrowly focused perspective of single-miR-based analysis, providing a low-dimensional, multi-miR-based analysis method able to characterize differences between leukemia types. The output is a network of small sets of miRs suitable for phenotyping, rather than a unique disease identifier, reflecting the inherent complexity of the disease and the redundancy of the blood/immune system.**

Low-dimensional representations of high-dimensional data can be obtained by means of standard statistical methods such as principal component analysis and singular value decomposition, although these approaches are not designed to separate different classes optimally [18]. Moreover, these and more recent methods (e.g. self-organizing maps [19] and multidimensional scaling [20,21]), achieve dimensional reduction by introducing more abstract descriptions of the system in terms of linear or non-linear combinations of the actual measurements, i.e. the number of gene or miR parameters that must be measured for phenotyping is not actually reduced. Furthermore, in the context of miR-based separation of phenotypes, the "principal components" that result from such low-dimensional procedures typically involve a large number of miRs, making a biological interpretation very difficult. The key question of "How many miRs (and which miR combinations) are needed to build a signature characteristic of a given acute leukemia?" remains largely unanswered by these methods.

In this context, we present a novel approach based on machine learning to reduce the number of parameters to manageable small sets of miRs, integrated with a method to represent the small sets using complex network representations. This procedure allows us to build and visualize multi-miR differential signatures of acute leukemias. Our work is focused on finding exhaustive lists of multidimensional miR groupings that characterize each of the 3 major types of acute leukemias (acute myeloid leukemia

[AML], B-precursor acute lymphoblastic leukemia [B-ALL], and T-cell acute lymphoblastic leukemia [T-ALL]), and visualizing these lists using network representations. Our approach is systematic (it searches all possible miR combinations), quantitative (it provides a measure of separation between leukemia types, so that one can compare quantitatively different miR groupings), scalable to larger group sizes (we analyze miR dyads and triads, which suffice to characterize all 3 acute leukemia types; however, the method can be applied to higher-dimensional miR groupings, e.g. miR tetrads), and unbiased (it treats every miR on an equal footing, without any a priori biological assumptions, although known biology could be used to bias the algorithm in a controlled fashion). We learn the multi-miR signatures based on data from well-characterized leukemia cell lines; then, we apply the learned patterns to primary patient leukemia samples, in order to determine the ability to predict and classify acute leukemias based on those signatures.

The experimental and analytic methods are described in the next Section. In the "Results and Discussion" Section, we describe some of the common approaches to analyze high-dimensional datasets, discuss their shortcomings and the need for a new approach, and present the new method to obtain AML-, B-ALL-, and T-ALL-specific multi-miR differential signatures. In the final Section, we summarize our findings.

## 2. Methods

### 2.1. Cell lines and primary cells

Precursor B-cell ALL, T-cell ALL, and AML cell lines (KASUMI1, K562, U937, KG1a, ML2, MO7e1, HL60 (AML), KOPN8, REH, RCHACV, KASUMI2, MHHCALL3, MHHCALL4, MUTZ5, NALM6, SUPB15 (B-ALL), JURKAT, CCRFCEM, MOLT3, MOLT16, KARPAS45 (T-ALL)) were purchased from American Type Cell Culture (ATCC, Manassas, VA, USA) and Deutsche Sammlung von Mikroorganismen und Zellkulturen GmbH (DSMZ, Braunschweig, Germany) and cultured according to the supplier's recommendations. Primary leukemia samples were obtained from cryopreserved primary samples or obtained from spleens of immunodeficient mice which had been transplanted with patient leukemia cells [22]. Primary CD34+ cells from mobilized peripheral blood from normal donors were purchased from the Hematopoietic Cell Processing Core (Fred Hutchinson Cancer Institute). Primary mature B lymphocytes, T lymphocytes, monocytes and granulocytes were purified from peripheral blood obtained from healthy adult human volunteers after informed consent following an Institutional Review Board-approved protocol (University of Maryland School of Medicine).

### 2.2. Isolation of RNA and miR profiling

Total RNA was extracted using miRNeasy Midi Kits (Qiagen, Valencia, CA) according to the manufacturer's guidelines. The quality of the isolated RNA was assessed using an Agilent 2100 Bioanalyzer. miR levels were measured using microarray analysis; ~500 ng RNA per sample was hybridized to each GeneChip miRNA 2.0 Arrray (Affymetrix,

Santa Clara, CA) which contained 46,228 probes for the miRs of 71 organisms. RNA labeling, hybridization to arrays, washing, staining, and signal amplification were performed according to manufacturer's instructions in the core facilities of the University of Maryland School of Medicine.

### 2.3. qRT-PCR Analysis

RNA was isolated from cells suspended in Triazol using miRNeasy kits (Qiagen, Valencia, CA) according to the manufacturer's protocol. Levels of mature miRs were measured in a 7900 Real-Time PCR System (Applied Biosystems) using TaqMan qRT-PCR (Applied Biosystems). U18 was used as an endogenous housekeeping gene control due to its consistent expression in acute leukemia cell lines. miR expression in every cell line was represented as Ct values, i.e. cycle number at which fluorescent signal crosses the pre-set threshold.

### 2.4. Data Analysis

Analysis of miR expression was performed using Partek Genomic Suite (Partek Inc., St. Louis, MO). The miR array expression data was normalized by means of robust multi-chip average (RMA) method, generating $\log_2$ transformed intensity values. Using this method we assessed 847 human miRs. In order to exclude the miRs that significantly and consistently express below detection levels, we further filtered based on miR expression with mean intensity ≥1 and ≤14, yielding 370 miRs [23,24]. We performed SVD dimensional reduction analysis by implementing the function 'svd' in R, which relies on the package MASS [25]. We performed Support Vector Machine analysis by implementing the function 'ksvm' in R, which is included in the package 'kernlab' [26]. We performed a pre-selection of Top miRs for the triad analysis using the Comparative Marker Selection and Comparative Marker Selection Viewer modules from the GenePattern Genomics Analysis suite [27]. Complex network images were produced using Cytoscape [28]. Our analysis framework was integrated by custom-written scripts in R, pro Fit, and Perl.

## 3. Results and Discussion

### 3.1. The importance of low-dimensional, miR-based signatures of acute leukemias

MiR expression was measured on AML, B-ALL, and T-ALL cell lines and primary patient samples. Initially, a set of 847 human miRs was measured for each sample; however, here and throughout, we focus on a subset of 370 miRs obtained after filtering out miRs with consistently low or absent expression (more details are provided in the Methods Section). For each cell line, 3 or 4 replicate measurements were taken. We tested the robustness of the replicate datasets by calculating Euclidean distances in the multidimensional miR-measurement space. We found that each replicate lay closer to the centroid of the remaining replicates from the same cell line, relative to its distance to the centroids of each of the other cell lines. All replicates passed this robustness test

(see Supporting Information Text S1 and Supporting Information Figures S1-S4 for more details).

Since samples are described by sets of 370 miR expression values, they can be represented as dots in a 370-dimensional space, where each axis in this high-dimensional representation corresponds to one specific miR. Naturally, such a high-dimensional representation cannot be visualized and, moreover, is of little help to convey the essential information oftentimes needed: Which and how many miRs need to be measured in order to fully distinguish one leukemia type from another? How accurately can these miR signatures, learned from leukemia cell lines, be used to classify various types of primary leukemia patient samples?

Principal Component Analysis (PCA) and Singular Value Decomposition (SVD) are standard methods for dimensional reduction, which select the directions of largest spread of the data and allow a low-dimensional representation that can be more easily handled and visualized. In the case where principal components are calculated from the covariance matrix, there is a direct relation between PCA and SVD: singular vectors from the SVD analysis then agree with the principal components from the PCA analysis, and the squared singular values become proportional to the variances of the principal components [29]. These techniques for dimensional reduction are powerful and have been used widely for analysis of microarray data [18,30-32]. Fig. 1(a) shows a PCA/SVD 2D representation of miR expression from AML (red), B-ALL (blue), and T-ALL (green) cell lines, where different replicate measurements from the same cell line are shown with the same symbol. However, in order to find a multi-miR signature that could be used to classify different leukemia subtypes, the PCA/SVD method suffers from limitations and shortcomings that require a different approach.

The main limitation of the PCA/SVD method is that the eigenmodes (such as modes 1 and 2 in Fig. 1(a)) are linear combinations of all the miRs in the original data. Thus, it is far from straightforward to interpret these leading characteristic modes in terms of specific miRs potentially useful for diagnosis or as targets for treatment. For instance, Fig. 1(b) shows the expansion coefficients of modes 1 and 2 in terms of the top 25 miRs that contribute to each mode. Clearly, many miRs contribute significantly to the leading eigenmodes; for instance, the number 25 miR contributing to mode 1 (namely, miR-93*) has a coefficient whose absolute value is 78% of the coefficient associated with the number 1 contributing miR (namely, miR-92a). Therefore, mode 1 is associated with many miRs; similar observations can be made on the relative contributions of the leading miRs in mode 2. In the case of other methods such as self-organizing maps [19,33], isomap [34], local embedding [35], and multidimensional scaling [20,21], the dimensionality-reduction process introduces non-linear transformations of the instance space, which suffers from similar limitations. Furthermore, although the PCA/SVD method is designed to find the directions of largest spread, it is not optimized to find the directions of largest separation between different data classes. Usually, the PCA/SVD low dimensional representation is informative of distinct data clusters (e.g. AML cell lines form an extended cluster of points on the lower region of Fig. 1(a), while B-ALL and T-ALL cell lines appear mostly on the upper region), but the PCA/SVD method is not designed to achieve class separation. Moreover, the separation of data classes may require dimensions higher than 3D, for which a visualization using PCA/SVD would not be possible.

In Fig. 1(c), miRs expressed in patients diagnosed clinically with AML (red), B-ALL (blue), and T-ALL (green) are represented in 2D using the same 2 modes from Fig. 1(a). The spatial distribution of patient samples by leukemia type is similar to that observed for cell lines, although somewhat more spread out and overlapping. Gene expression profiling has shown consistent differences between primary clinical samples of human leukemias and their model cell lines, where the former appeared to overexpress genes related to the immune response, while the latter showed higher expression of genes related to macromolecular metabolism [36]. However, after eliminating other cell-line-specific signatures (proliferation, immortalization, etc.), additional molecular profiling studies have shown that characteristic genomic aberrations were conserved across leukemic clinical and cell line samples, and these signatures were quite robust, as expression data from cell lines correlated highly with previously published data [37]. Therefore, our working hypothesis is that differential miR signatures learned from cell line data can be meaningfully used as predictors of clinical specimen classification.

By comparing, one miR at a time, the distribution of miR expression values from different acute leukemia types, we observe that the distributions overlap with each other: That is, we cannot fully separate e.g. AML samples from B-ALL and T-ALL samples based only on the expression level of any one miR. Thus, we test whether taking two or more miRs simultaneously into account achieves separation. Using two or more miRs, the separation between samples of different classes can be assessed by means of standard machine learning methods. Machine learning approaches such as Support Vector Machines (SVMs) have been tailored as methods of gene selection for cancer classification [38]. In this work we use a Support Vector Machine with a linear kernel, which is equivalent to the machine learning method known as the perceptron [39,40], in order to avoid overfitting and, moreover, to obtain a straightforward interpretation of the machine-learned parameters in terms of the original measurements [41,42]. Since this approach relies heavily on statistics, we used all cell line replicate measurements as classification instances in the learning phase (24 samples from AML, 36 from B-ALL, and 20 from T-ALL cell lines). As shown in the outlier analysis (see Supporting Information Text S1 and Supporting Information Figures S1-S4), all replicates passed the robustness test and are considered faithful representatives of their parent cell lines.

### 3.2. AML-centric miR-dyad networks
Let us first consider AML-centric multi-miR signatures. For each pair of miRs, we perform an SVM analysis to determine whether samples are linearly separable based on the leukemia types (i.e. we determine whether a straight line boundary exists to fully separate the AML vs B-ALL cell lines, as well as the AML vs T-ALL cell lines). If the leukemia-type-specific samples are not separable, we disregard that given miR pair, whereas if the samples are separable, we label that given pair as a "miR dyad" and we compute the gap distance (classifier margin) between the different classes. We verify the robustness of the linear separation by applying leave-one-out cross-validation. By repeating this procedure over the 68,265 pairwise combinations from the initial set of 370 miRs, we generate a so-called "AML-centric dyad network". Let us point out that other machine learning implementations are possible, e.g. soft-margin and non-linear classifiers, which would extend the current application to scenarios that tolerate class overlaps due to noise in the datasets. Those model extensions are beyond the scope of the present paper. The constraint to strictly linearly separable classes leads to a straightforward implementation that, despite its relative simplicity, is able to capture the redundancy of multi-miR signatures.

Fig. 2(a) shows the AML-centric miR-dyad network obtained with cell line data, where any connected pair of miRs achieves separation between AML and B-ALL (blue links) as well as between AML and T-ALL (red links) cell lines. The link width reflects the gap distance between different classes (i.e. the width of the boundary determined by the support vectors of each class), which is a measure of the goodness of the class separation. This network contains 15 nodes (each one corresponding to a different miR) and 18 link pairs (each blue/red link pair corresponding to a different miR dyad). The miR-dyad network of Fig. 2(a) is a novel type of low-dimensional representation that conveys miR-specific information on how to optimally separate different leukemia types, unlike the standard PCA/SVD representation of Fig. 1(a).

Let us now validate our findings by analyzing the AML-centric networks in the context of the miR literature. A particularly well-studied set of miRs that stands out in our analyses is that formed by miR-23a, miR-24, and miR-27a. miR-23a and miR-24 were reported to be enriched in CD34+ hematopoietic stem-progenitor cells (HSPCs) [43], and miR-24 was identified to have a well-defined role as a regulator of normal erythropoiesis via targeting of human activin receptor type1, ALK4 [44]. Moreover, miR-23a and miR-24 both act as tumor suppressors and are down-regulated in multiple cancers. miR-27a is co-localized in mammalian genomes with miR-23a and miR-24 [45]; these form the so-called "miR-23a cluster" that is coordinately transcribed and is under direct negative regulation by c-MYC [46]. All 3 miRs in this cluster were reported to be down-regulated in acute promyelocytic leukemia, colorectal cancer, prostate cancer, and oral squamous cell carcinoma [47-49]. Furthermore, it was recently shown that, by interacting with and regulating the pro-apoptotic members of the 14-3-3θ gene family, miR-27a functions as a tumor suppressor in acute leukemia [50]. The AML-centric miR-dyad network displays all 3 miRs from the miR-23a cluster. Indeed, miR-27, with 9 links, appears as the most connected miR in the network, whereas miR-24 has 3 links and miR-23a one connection. It should be noted that these miRs are not connected with each other; instead, they appear to share neighbors. This result can be understood from the fact that two miRs are not connected based on their similarity but rather based on the ability to fully separate AML from B-ALL and T-ALL. Thus, one might expect that coordinately expressed miRs share too much redundant information to make them contribute independently to a suitable signature to fully characterize AML. Another interesting observation is that all 3 miRs from the miR-23a cluster appear connected to miR-145, which is a well characterized miR with tumor suppressor roles reported in a variety of cancers [51-53]. Moreover, expression of miR-145 and all its linked miRs in the network (namely, miR-23a, 23a*, 24, 27a, 27a* and 130b) has been shown to be involved in proliferation and differentiation of hematopoietic cells and to be altered during leukemogenesis [54,55].

Fig. 2(b) shows the number of links associated with each miR in the network. We observe that the most connected node is miR-27a (with 9 connections), followed by miR-145 (with 6 connections); indeed, 14 out of the 18 miR dyads that characterize AML vs both B-ALL and T-ALL involve either miR-27a or miR-145. Fig. 3(a) shows the distribution of cell line samples for miR-145 vs miR-27a, where, as expected, AML samples (red symbols) appear well separated from both B-ALL samples (blue symbols) and T-ALL samples (green symbols). As in Fig. 1, different replicate measurements from the same cell line are shown with the same symbol. In Fig. 3(b), we show miR-145 vs miR-27a for patient samples; the pattern of spatial distribution is similar to that for cell lines, although with a somewhat larger spread. Indeed, a slight overlap between AML and B-ALL is observed in the $27a^+/145^-$ region (bottom right corner of Fig. 3(b)). As suggested by Figs. 3(a)-(b), we should expect cell lines to be very good predictors of clinical sample separation in miR space, although some miR dyads may fail to correctly classify a few patient samples, which may actually be at a diagnostic borderline between e.g. AML and B-ALL [56].

Fig. 4(a) shows the AML vs B-ALL prediction of clinical samples using AML-centric miR dyads. The horizontal axis corresponds to 18 samples from patients diagnosed with AML and 29 samples from patients diagnosed with B-ALL, while the vertical axis shows the 18 miR dyads from the AML-centric network (Fig. 2(a)). Correctly classified samples are shown in green, while incorrectly classified samples are displayed in red. The overall prediction performance is very good, with 94% of patients correctly classified when considering all of the miR dyads found in cell lines. This matrix representation highlights evidence that most failed predictions arise from a few borderline clinical samples (namely, patients AML_3, AML_6, AML_8, B-ALL_1, and B-ALL_3) which represent only 11% of the samples but concentrate 76% of the total prediction failures. Similarly, Fig. 4(b) shows the AML vs T-ALL prediction of clinical samples using AML-centric miR dyads. In this case, the prediction performance is excellent (99.5% correct).

### 3.3. B-ALL- and T-ALL-centric miR-triad networks
In the case of B-ALL-centric networks, there are no miR dyads that fully separate all available B-ALL cell line samples from all AML and T-ALL cell line samples. The next logical step is to investigate miR triads. However, instead of searching through all possible triad combinations, which is very computationally intensive (e.g. N=370 miRs would require $8.4 \times 10^6$ iterations), here we lay out a framework that is scalable to situations where analyzing all possible combinations is outright unfeasible, such as, for instance, the case of larger miR datasets (e.g. $1.7 \times 10^8$ triad combinations out of N=1000 miRs) or the investigation of higher-order groupings (e.g. $7.7 \times 10^8$ tetrad combinations out of N=370 miRs). This scalable framework combines our machine-learning-based network approach with the class prediction method developed in Ref. [57]. By using an implementation of the latter made available by the Broad Institute's GenePattern Genomics Analysis platform [27], each miR is assigned a score that measures its significance in a binary classification problem (e.g. classifying B-ALL vs AML samples). This set of scores is renormalized by subtracting the mean and dividing by the standard deviation of the score distribution (i.e. so that the distribution of renormalized scores has zero mean and unit variance). In this way, we obtain two sets of scores, one corresponding to the B-ALL vs AML case, the other one to the B-ALL vs T-ALL case. We combine them by choosing, for each miR, the lowest score. Notice that, since our focus is on achieving full separation of B-ALL from AML and T-ALL, selecting the lowest score follows a "weakest-link" rationale that is more appropriate than combining the scores by other procedures, e.g. root-mean-square averages. This set of combined normalized scores (which we call "GP scores") is then rank-ordered from most to least significant. By selecting the resulting Top-50 miRs in the rank-ordered GP score list, we effectively reduce the original dataset to a smaller one, in which the search for relevant triad combinations is more tractable. As pointed out above, this framework would also be useful for searches of higher-order groupings, such as e.g. miR tetrads.

Based on this Top-50 GP miR list, we apply the machine-learning network method as described above, which leads to the B-ALL-centric triad network ensemble shown in Fig. 5(a). This ensemble consists of several networks, each associated with a hub node (shown in red). Triads are formed by adding the hub node to each pair of connected nodes in the network; hub nodes from other networks in the ensemble are shown in dark yellow. The largest network in the ensemble is associated with the hub node representing miR-708; a triad formed by adding miR-708 to any pair of connected nodes in this network (for instance, the triad miR-708/503/652) achieves full separation between all B-ALL and AML cell line samples, as well as between all B-ALL and T-ALL cell line samples. Notice that e.g. miR-503 and miR-652 appear dark yellow in the network associated with hub node miR-708, because these nodes are, in turn, hub nodes of other networks in the ensemble. This network ensemble contains 191 miR triads. Notice, however, that, due to the large number of triads, the gap size between leukemia classes cannot be easily represented as link widths (similarly to the AML-centric dyadic network in Fig. 2(a)). Instead, we provide the full list of class separation gap sizes as Supporting Information Dataset S1.

An additional important measure of the role of a given miR is provided by the number of triads in which the miR appears. Fig. 5(b) shows, for each node in the network ensemble, the number of triads in which the node appears. As mentioned, the most connected node is miR-708, which appears in 70% of the triads (133 of the 191 triads in the network ensemble). Following the learning phase, in which we generated triad networks based on data from acute leukemia cell lines, we predicted the classification of primary clinical acute leukemia samples. Using all 191 triads from the network ensemble, we determined, as a function of the percentage of patients in the cohort (x axis), the number of triads that correctly predict a number of patients equal to or larger than x. This is shown (for the B-ALL vs AML case) by the red plot in Fig. 5(c). These prediction results are compared with average prediction outcomes for 1000 triad sets (of 191 triads per set) generated from miRs chosen by other methods, namely: triads from Top-50 miR lists obtained with GenePattern [27] (labeled "GP Top-50", blue plot), triads from Top-50 miR lists obtained with Support Vector Machine in 370-D (labeled "SVM Top-50", green plot), and triads randomly picked from the full list of 370 miRs (labeled "random", black plot). This comparison shows that the triad network ensemble outperforms other methods of miR triad selection in their ability to classify primary acute leukemia patient samples. For instance, 177 triads (out of 191) from the triad network ensemble are able to predict correctly at least 75% of the patients in the cohort. By using the "GP Top-50" method, 136 triads are able to predict correctly the classification of at least 75% of the patients, whereas the "SVM Top-50" method provides only 123 triads above the prediction threshold. By randomly selecting 191 miR triads, only 101 triads (on average) are able to predict correctly at least 75% of the patients; as expected, random selection has the worst performance. Analogously, Fig. 5(d) shows a comparison of prediction outcomes in the patient cohort for the B-ALL vs T-ALL case, where the triad network ensemble is clearly seen to outperform other methods of miR triad selection.

In Fig. 6 we show results pertaining to the T-ALL-centric triad network ensemble. The networks shown in Fig. 6(a) contain a total of 136 miR triads that are capable of fully separating all T-ALL cell line samples from both AML and B-ALL (see Supporting Information Dataset S2 for the full list of class separation gap sizes). The most connected node is miR-222, which appears in 55 miR triads (see Fig. 6(b)). These triads (learned by using cell lines) show excellent prediction outcomes to classify T-ALL from AML primary patient samples (Fig. 6(c)). Indeed, all 136 triads are able to predict correctly at least 75% of the patients, 129 triads predict 85% or more of the cohort, and 102 triads have a prediction outcome of 95% and above. The performances of other methods of triad selection are shown for comparison.

The miR-23a cluster (which, as mentioned above, includes miR-23a, 24, and 27a) also appears in the miR-triad network ensembles obtained for B-ALL (Fig. 5(a)) and T-ALL (Fig. 6(a)), an observation that emphasizes its key role in leukemia. As with the AML-centric network, no direct links are observed between miRs in the miR-23a cluster, but they appear to share many neighbors. For instance, the B-ALL network with hub node miR-146a shows miRs 23a, 24, 27a, and others connected to miR-455-3p. This means that triads of the form (146a, 455-3p, X) are able to fully characterize B-ALL, where X is any miR from the miR-23a cluster. Similarly, the B-ALL network with hub node 455-3p shows miRs 23a, 24, 27a, and 193b connected to miR-182, which points to triads of the form (455-3p, 182, X). In addition to the miR-23a cluster, it is interesting to point out other miRs that were shown to play a crucial role in normal and malignant B-lymphoid development, e.g. miR-34a, miR-18a, and miR-151 [54,55,58], which also appear in the B-ALL-centric triad network. Notice e.g. miR-151-5p, which appears on 51 triads and is the second-most connected miR in the B-ALL network, while it is absent altogether from the AML and T-ALL-centric networks.

Multi-miR networks may be used as exploratory tools to identify promising miRs and guide further experimentation. Indeed, the results for B-ALL (Fig. 5) provide strong evidence of the remarkable significance of miR-708, as a biomarker and potentially for some function. We performed quantitative reverse transcription polymerase chain reaction (qRT-PCR) analysis of relative miR-708 expression levels across all studied cell lines to validate the miR microarray results. Figure 7 shows the threshold cycle (Ct value) for each cell line, i.e. the number of cycles needed for the measured fluorescence to exceed the threshold intensity (see the Methods Section for more details). Indeed, miR-708 is up-regulated in most B-ALL cell lines compared with T-ALL and AML, suggesting that miR-708 might serve as a biomarker for B-ALL.

### 3.4. Driving phenotype changes by miR modulation: The need for a multi-miR approach

The association of miR dysfunction with disease phenotypes has given rise to the idea that selective modulation of miRs could affect the disease [59]. The feasibility of this therapeutic rationale has indeed been confirmed by the striking demonstration that inhibition of miR-122 reduces viral load in chronically HCV-infected chimpanzees [60]. Attempts to drive phenotype changes with single-miR modulation approaches, however, may be misleading, as illustrated by the example of Figure 8. In Figure 8(a), we compare the expression levels of miR-23a in CD34+ cells, monocytes and granulocytes from healthy donor samples (left box) with AML patient samples (right box). AML expression levels of miR-23a are significantly lower than those of healthy donor cells ($p=0.001$). In order to push the miR-23 expression level towards the level in normal cells, a miR modulation experiment would increase the expression of miR-23a in the AML samples (dotted arrow). However, miR-23a by itself is not able to separate (and, hence, fully distinguish) AML samples from those of healthy cell counterparts. Figure 8(b) shows that miR-23a and miR-24, taken together as a 2D signature, are able to achieve separation. The boundary between the healthy and diseased regions is shown as a solid line, while a solid arrow shows the direction perpendicular to the boundary. By increasing miR-23a, however, an AML sample (such as the one marked with a circle) would move along Trajectory I and remain within the diseased region of this 2D parameter space. The optimal direction to cross the boundary towards the healthy phenotypes is indicated by Trajectory II, which involves decreasing miR-23a and increasing miR-24 according a 1:3 ratio. Thus, only by considering a multi-miR signature do we observe the optimal trajectory in which we would rightly expect a phenotype change. This example illustrates that expectations based on 1D (i.e. single-miR) approaches may be misleading and shows the need of a multi-miR approach to properly design experiments such as selective miR loss- or gain-of-function analysis [61].

## 4. Conclusions

In this work, we developed a novel dimensional-reduction approach based on machine learning and complex network representations, which allowed us to build multi-miR differential signatures of the 3 major types of acute leukemias. Starting with AML, we found 18 miR dyads that fully separate AML from B-ALL and T-ALL cell lines. We visualized our findings through what we call a miR-dyad network that is centered around one specific phenotype, in this case AML. Starting with B-ALL or T-ALL, miR dyads did not suffice to fully differentiate these leukemias from the other subtypes. Therefore, we went on to implement a scalable framework that first preselects the most promising miRs based on single-miR analysis, followed by a machine learning approach designed to find differential signatures of miR triads. Based on those findings, we computed and visualized B-ALL and T-ALL-centric miR-triad network ensembles.

Multi-miR networks provide information on differential molecular signatures characteristic of different acute leukemia types. This information can be used in different ways. In the context of studies specific to one particular miR, the multi-miR network shows which other miRs might be associated with the given miR of interest, leading to dyads or triads that together form a distinctive signature. For instance, researchers investigating miR-886-3p might use the information from the AML-centric network to incorporate miR-27a to their investigations (e.g. by studying the response of leukemia cell lines to varying concentrations of exogenous miR-886-3p in combination with miR-27a).

Alternatively, multi-miR networks can be used as exploratory tools to identify interesting miRs and stimulate further experimentation. Indeed, after discovering the remarkable features of miR-708 in the B-ALL network, we performed qRT-PCR analyses across all studied cell lines, which confirmed that miR-708 was highly and specifically expressed in most B-ALLs, hence suggesting that miR-708 might serve as a biomarker for B-ALL. These results, indeed, agree with earlier reports on aberrant miR expression profiles in childhood ALL [15,62,63]. Schotte et al. [15] report that miR-708 was 250-6500-fold more highly expressed in TEL-AML1, BCR-ABL, E2A-PBX1, hyperdiploid and other B-ALL cases than in MLL-rearranged and T-ALL cases ($P<0.01$). Han et al. [62] investigated the association of dysregulated miRs with childhood ALL clinical outcome and found that miR-708 was the most upregulated miR in primary samples obtained at the time of leukemia relapse. miR-708 was also found to be associated with in vivo glucocorticoid therapy response and with disease risk stratification. Furthermore, Li et al. [63] reported significant upregulation of miR-708 in B-ALL samples ($P < 0.05$) and elevated expression of miR-708 in high-risk B-ALL compared to standard- and intermediate-risk cases. CNTFR, NNAT, and GNG12 were identified as targets of miR-708 [63].

Finally, we discussed the importance of multi-miR approaches in the context of miR modulation experimental design. By comparing AML samples with normal cell counterparts (CD34+, monocytes and granulocytes) from healthy donor samples, we showed that expectations based on miR-23a alone were misleading. By incorporating miR-24 to form a 2D miR signature that fully separates AMLs from normal, we found the optimal trajectory to cross the boundary from the AML region to the healthy region, which involved decreasing miR-23a, instead of increasing it (as was initially suggested by the single-miR analysis based solely on miR-23a).

To the best of our knowledge, this kind of experimental multi-miR approach, informed by a systematic and unbiased method such as the one presented in this work, has not been reported previously. We believe that identifying a large set of possible pairs or triads of phenotype relevant miRs and representing them in a visual way as a network has strong potential for discovery research, as well as for clinical diagnosis and prognosis. Indeed, although differential diagnosis between AML, B-ALL, and T-ALL is, in most cases, successfully achieved by classic histology, flow cytometry biomarkers and/or immunohistochemistry, our results suggest that multi-miR panels have the potential to be useful diagnosis tools for clinically ambiguous leukemias [13,14]. Furthermore, this approach can be applied to comparisons of a leukemia and its healthy counterparts, as briefly discussed in Figure 8, and to cohorts with different outcomes (e.g. cured vs relapsed patients), thus providing a window into prognosis and the quantitative evaluation of disease progression under different treatment strategies.

Finally, let us point out that, although this work has focused on the role of complex networks in combination with machine learning as dimensional-reduction and visualization tools for data analysis, our framework naturally suggests follow-up studies to quantitatively characterize the structure and dynamics of multi-miR networks through topological measures such as degree and path length distributions, clustering, modularity, assortative mixing, etc [8,64], which may provide valuable systems-level insight into molecular processes in health and disease.

## Acknowledgments

This work was supported in part by grants from the National Foundation for Cancer Research, the National Institutes of Health (P01CA70970 and T32CA154274) and the Maryland Stem Cell Research Foundation/TEDCO (2007-MSCRFII-0114 and 2010-MSCrFII-0065-00). The content is solely the responsibility of the authors and does not necessarily represent the official views of the National Cancer Institute or the National Institutes of Health.

## Disclosure/conflict of interest

The authors have no disclosures to report.

### Ethics Statement

This investigation was approved by Institutional Review Board of the University of Maryland Baltimore and conducted according to the principles expressed in the Declaration of Helsinki.

### Data Repository

The data discussed in this publication have been deposited in NCBI's Gene Expression Omnibus and are accessible through GEO Series accession number GSE51908 (http://www.ncbi.nlm.nih.gov/geo/query/acc.cgi?acc=GSE51908).

### Supporting Information

**Supporting Information Text S1:** Outlier analysis of cell-line replicates: centroid distances in multidimensional miR-measurement space.
**Supporting Information Figure S1:** Schematic representation of the centroid-based method for outlier analysis.
**Supporting Information Figure S2:** Outlier analysis of AML cell line data: 24 replicates from 7 cell lines.
**Supporting Information Figure S3:** Outlier analysis of B-ALL cell line data: 36 replicates from 9 cell lines.
**Supporting Information Figure S4:** Outlier analysis of T-ALL cell line data: 20 replicates from 5 cell lines.
**Supporting Information Dataset S1:** List of miR triads in the B-ALL-centric network, which includes normalized gap sizes of separation for B-ALL vs AML and B-ALL vs T-ALL [available upon request to corresponding author].
**Supporting Information Dataset S2:** List of miR triads in the T-ALL-centric network, which includes normalized gap sizes of separation for T-ALL vs AML and T-ALL vs B-ALL [available upon request to corresponding author].

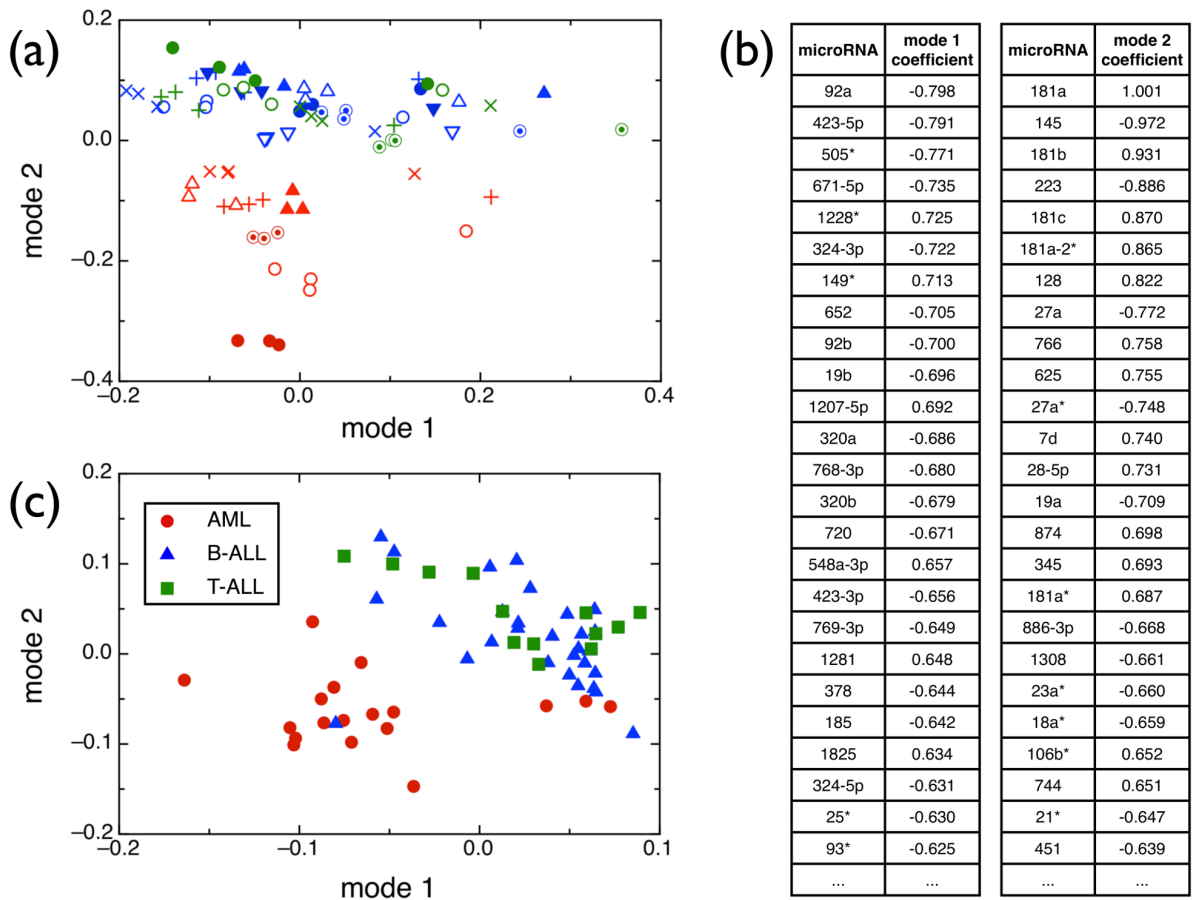

**CAPTION TO FIGURE 1:**

**Standard methods for dimensional reduction: Principal Component Analysis (PCA)/Singular Value Decomposition (SVD). (a)** 2D representation of AML (red), B-ALL (blue), and T-ALL (green) miR expression from **cell lines**, where different replicate measurements from the same cell line are shown with the same symbol. The leading eigenmodes capture the directions of largest spread of the data. **(b)** Each mode in the PCA/SVD decomposition is a linear combination of all measured miRs. The rank-ordered list of the top 25 miRs that contribute to mode 1 is shown on the left-hand side table, whereas the top 25 miRs that contribute to mode 2 are shown on the right-hand side. Coefficients are normalized so that the sum of absolute values (over all the 370 measured miRs) equals 100. **(c)** Using the same 2 modes, miR samples from **patients** diagnosed with AML (red), B-ALL (blue), and T-ALL (green) are represented in 2D.

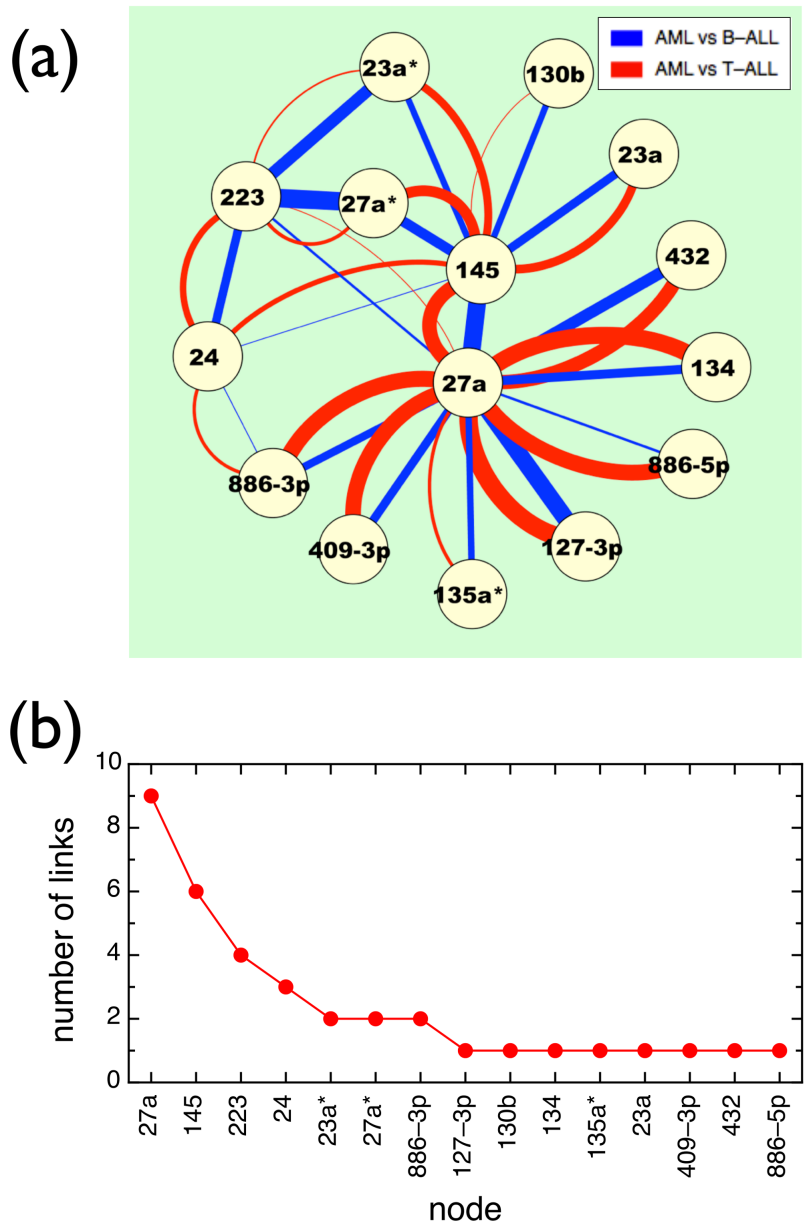

**CAPTION TO FIGURE 2:**

**AML-centric miR-dyad network obtained with cell line data. (a)** Any connected pair of miRs achieves separation between AML and B-ALL (blue links) as well as between AML and T-ALL (red links). The link width indicates the gap distance between different classes. This network contains 18 miR dyads. **(b)** Degree distribution of the AML dyad network, which shows the number of links associated with each miR.

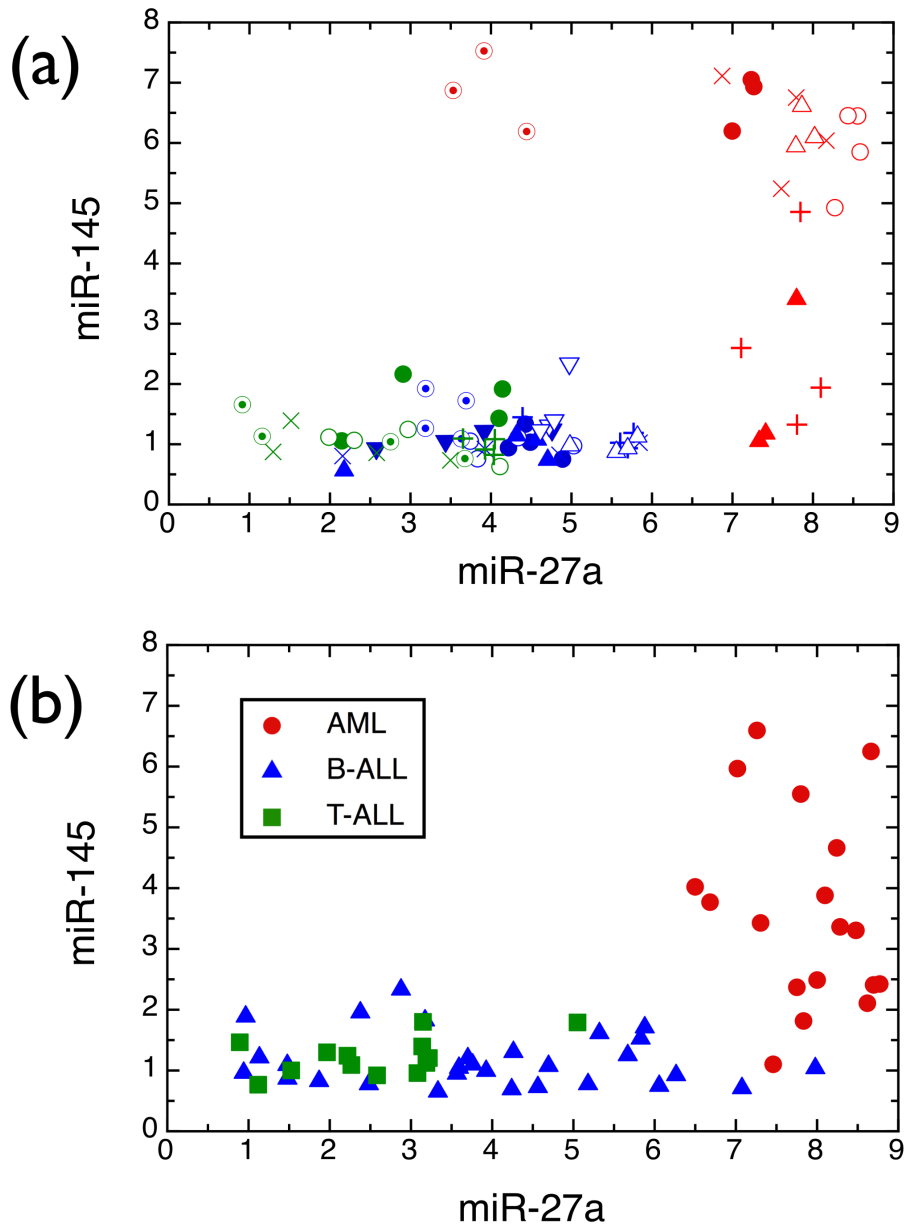

**CAPTION TO FIGURE 3:**

**Distribution of samples in a 2D miR space: miR-145 vs miR-27a. (a)** For cell lines, AML (red) samples are well separated from both B-ALL (blue) and T-ALL (green) samples. As in Fig. 1, different replicate measurements from the same cell line are shown with the same symbol. **(b)** Patient data are more inhomogeneous: a slight overlap between AML and B-ALL is observed in the $27a^+/145^-$ region (bottom right corner).

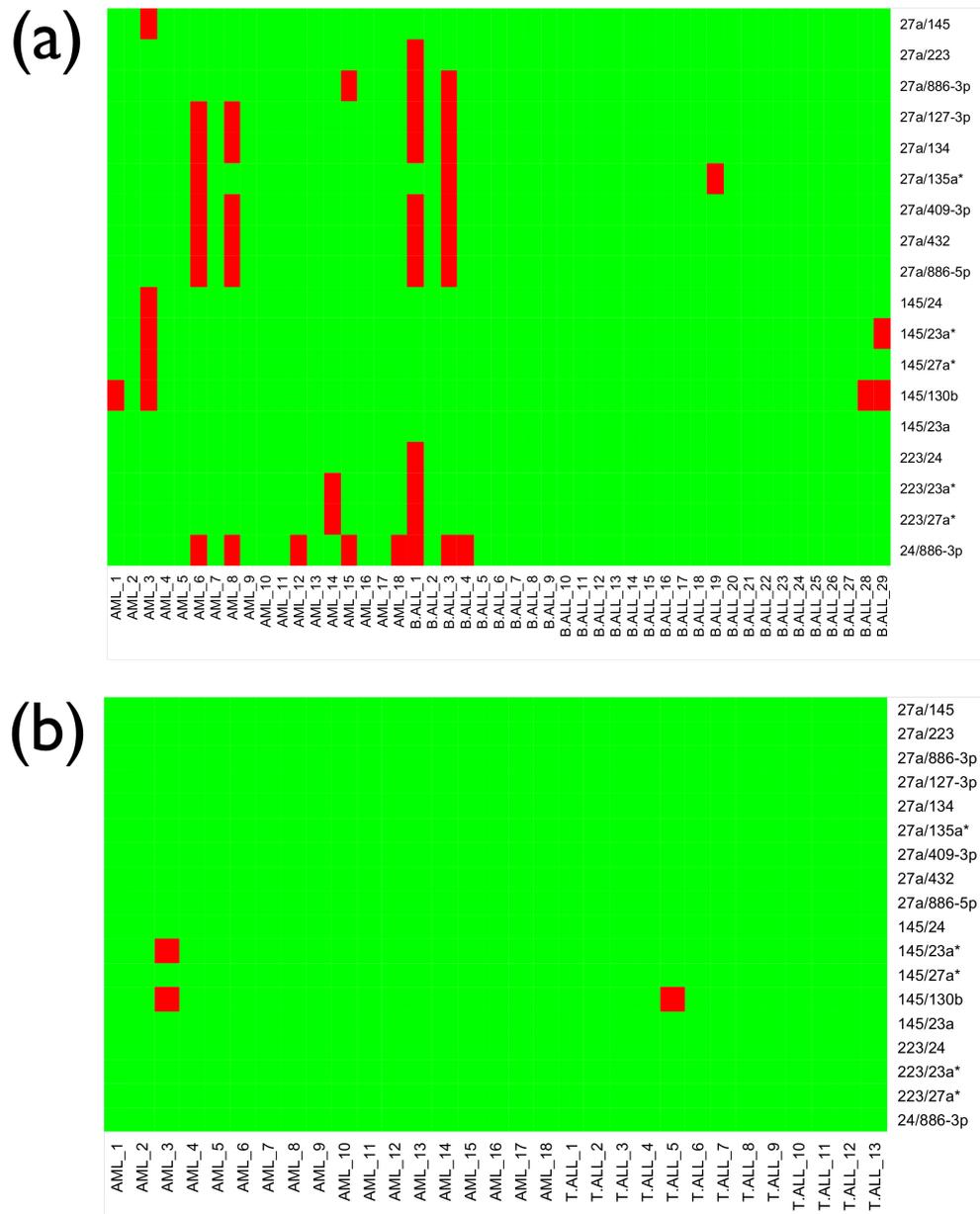

**CAPTION TO FIGURE 4:**

**Prediction of patient samples using AML-centric miR dyads.** The horizontal axes show patient samples, whereas the vertical axes correspond to the 18 miR dyads from the AML-centric network. Correctly classified samples are shown in green, incorrectly classified samples are displayed in red. **(a) AML vs B-ALL** case: Classification of 18 AML and 29 B-ALL patients. **(b) AML vs T-ALL** case: Classification of 18 AML vs 13 T-ALL patients.

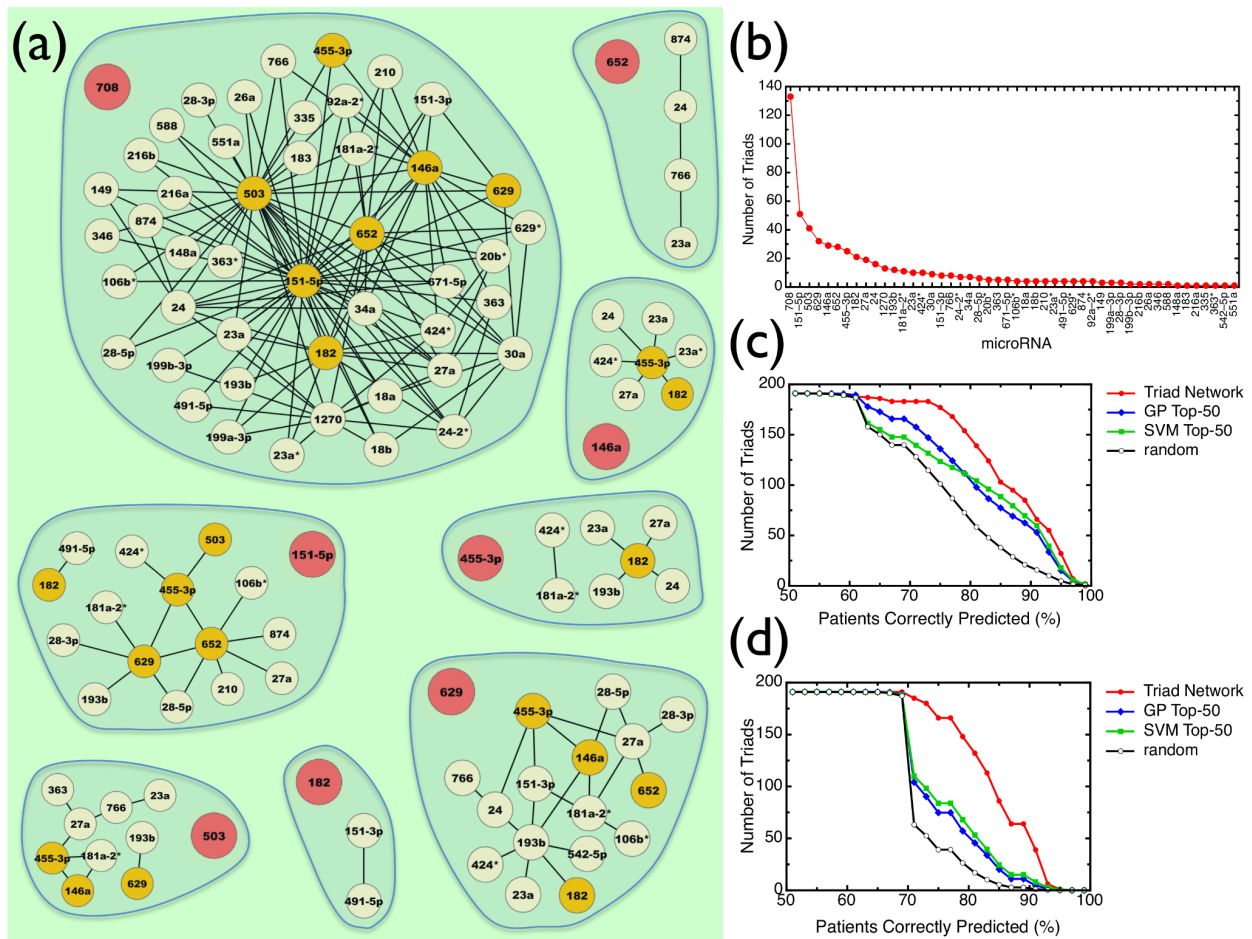

**CAPTION TO FIGURE 5:**

**B-ALL-centric network ensemble of miR triads.** **(a)** Networks obtained with cell line data. Each network of pairwise connections has a hub node (in red); triads are formed by adding the hub node to each pair of connected nodes. Hub nodes from other networks in the ensemble are shown in dark yellow. **(b)** Number of triads in which each node appears. The most connected node is miR-708, which appears in 133 out of 191 triads. **(c)** As a function of the percentage of patients in the cohort (x axis), these plots show (for the **B-ALL vs AML** case) the number of triads that correctly predict a number of patients equal or larger than x. The prediction results using all triads from the triad network (red plot) are compared with average prediction outcomes for 1000 triad sets (of 191 triads per set) generated from miRs chosen by other methods, namely: triads from Top-50 miR lists obtained with GenePattern [27] (labeled "GP Top-50", blue plot), triads from Top-50 miR lists obtained with Support Vector Machine in 370-D (labeled "SVM Top-50", green plot), and triads randomly picked from the full list of 370 miRs (labeled "random", black plot). **(d)** Analogously to (c), prediction outcomes on the patient cohort for the **B-ALL vs T-ALL** case, where the triad network is compared to other methods of selecting miR triads.

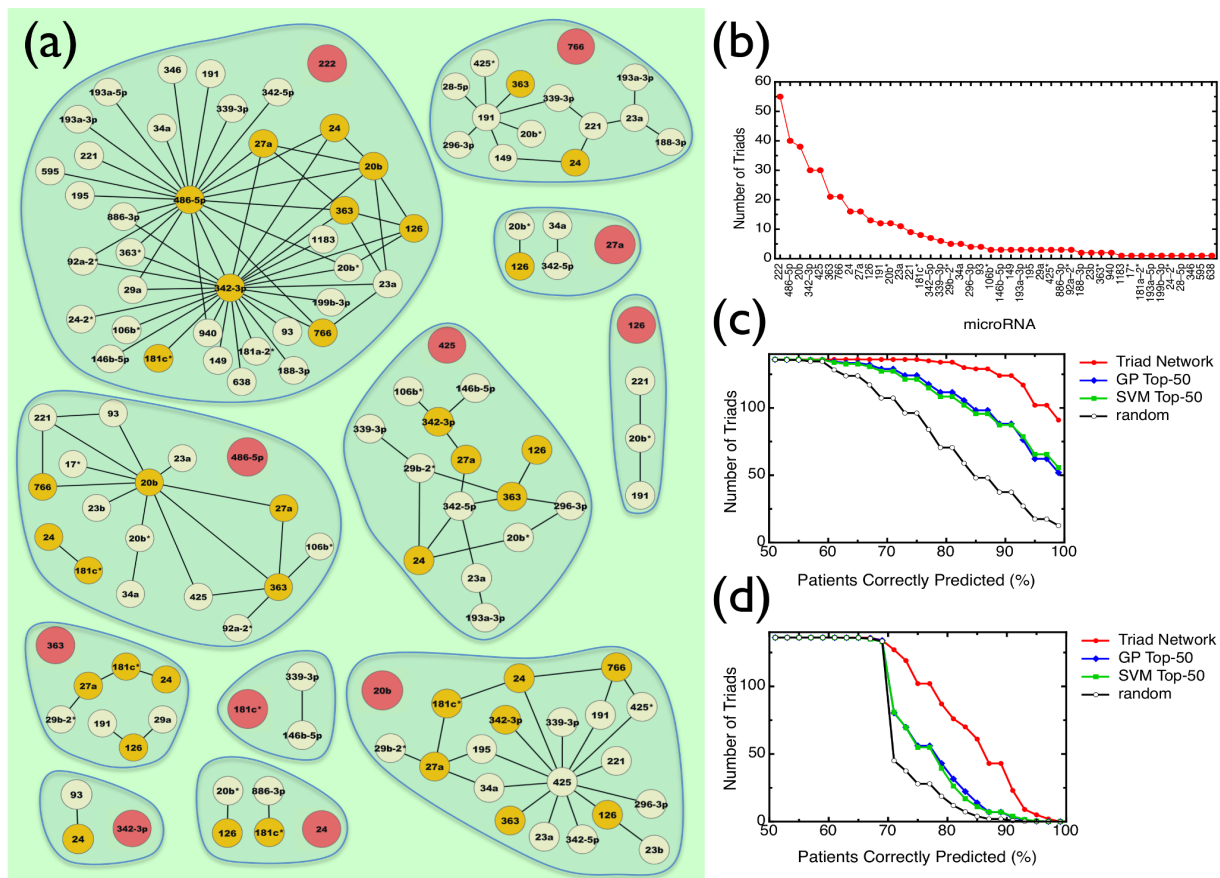

**CAPTION TO FIGURE 6:**

**T-ALL-centric network ensemble of miR triads. (a)** Networks obtained with cell line data. Each network of pairwise connections has a hub node (in red); triads are formed by adding the hub node to each pair of connected nodes. Hub nodes from other networks in the ensemble are shown in dark yellow. **(b)** Number of triads in which each node appears out of a total of 136 triads. **(c)** As a function of the percentage of patients in the cohort (x axis), these plots show (for the **T-ALL vs AML** case) the number of triads that correctly predict a number of patients equal or larger than x. The prediction results using all triads from the triad network (red plot) are compared with average prediction outcomes for 1000 triad sets (of 136 triads per set) generated from miRs chosen by other methods, namely: triads from Top-50 miR lists obtained with GenePattern [27] ("GP Top-50", blue plot), triads from Top-50 miR lists obtained with Support Vector Machine in 370-D ("SVM Top-50", green plot), and triads randomly picked from the full list of 370 miRs ("random", black plot). **(d)** Analogously to (c), prediction outcomes on the patient cohort for the **T-ALL vs B-ALL** case, where the triad network is compared to other methods of selecting miR triads.

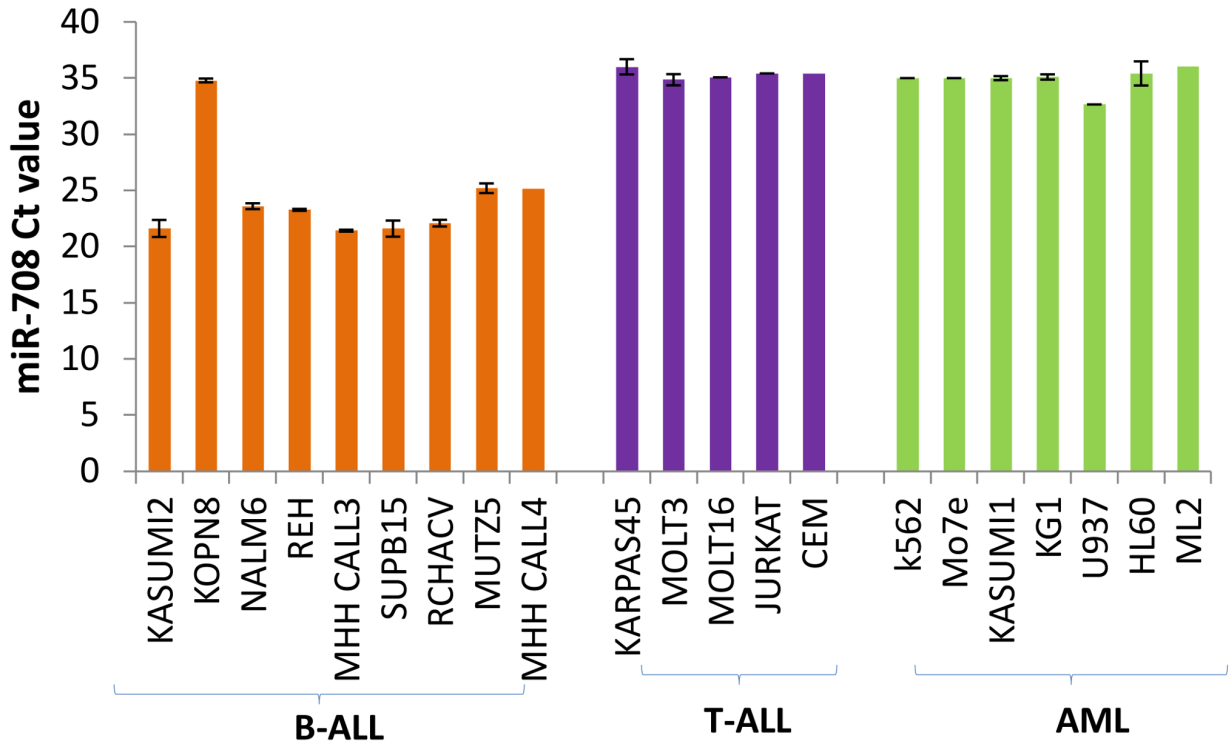

**CAPTION TO FIGURE 7:**

**qRT-PCR analysis of relative miR-708 expression levels in B-ALL, T-ALL, and AML cell lines.** Relative threshold cycles (Ct values), determined by the number of cycles needed for the measured fluorescence to exceed the threshold intensity, is shown for every cell line. Ct values show that miR-708 is up-regulated in most B-ALL cell lines, as compared to T-ALL and AML cell lines.

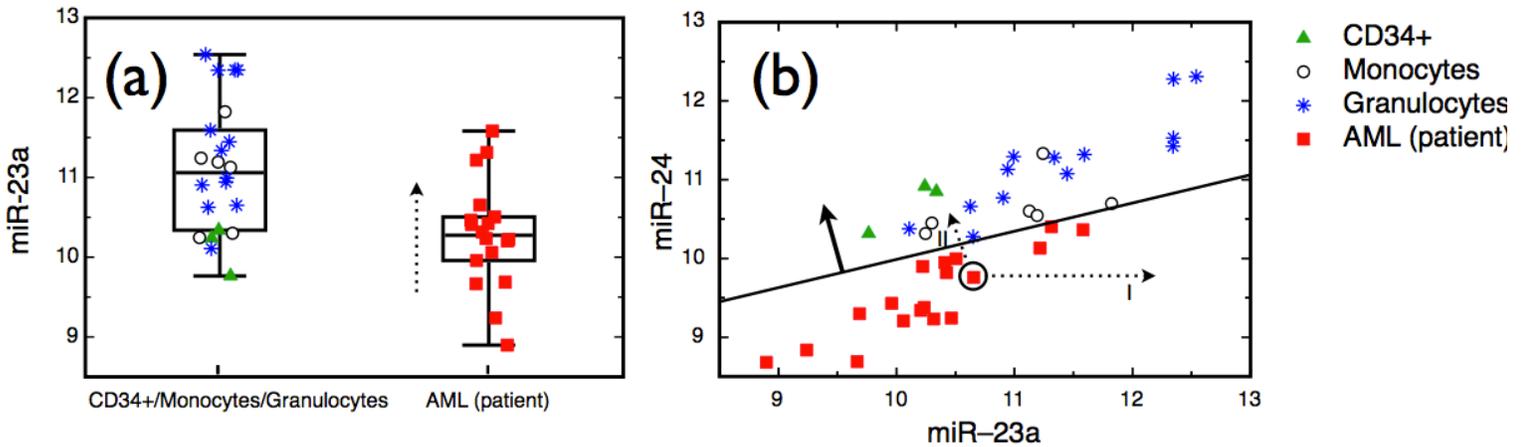

**CAPTION TO FIGURE 8:**

**Single- vs Multi-miR approaches to driving phenotype changes by miR modulation.** (a) Comparison between AML patient data (18 samples) and data for normal cell types (22 samples from healthy donors). The p-value is 0.001. The "1-miR approach" to modulate AML patient samples based on this information suggests to increase miR-23a (dotted arrow). Notice, however, that these populations overlap, meaning that miR-23a alone does not suffice to classify an unknown sample as healthy or AML. (b) The combination of miR-23a with miR-24 is able to fully classify these samples as normal or AML. The SVM boundary (solid line) finds the optimal class separation. By increasing miR-23a alone (Trajectory I), an AML sample remains in the diseased region. Trajectory II (which involves decreasing miR-23a as miR-24 is increased) shows the shortest path to traverse the boundary from the AML region to the healthy region.

## Supporting Information Text S1

**Outlier analysis of cell-line replicates: centroid distances in multidimensional miR-measurement space**

MiR expression was measured on 7 AML cell lines (HL60, K562, KASUMI1, KG1a, ML2, MO7e1, and U937), 9 B-ALL cell lines (KASUMI2, KOPN8, MHHCALL3, MHHCALL4, MUTZ5, NALM6, RCHACV, REH, and SUPB15), and 5 T-ALL cell lines (CCRF_CEM, JURKAT, KARPAS45, MOLT16, and MOLT3). Each cell line has been measured several times (3 to 4 replicates per cell line). In order to ensure that our machine-learning analysis is based on robust data, we developed the following robustness test to identify possible outliers.

Let us assume that, for a given disease class, $n_s$ samples have been measured from $n_{cl}$ different cell lines (e.g. for AML, $n_s$=24 samples have been measured from $n_{cl}$=7 different cell lines). Measured samples are labeled by subindex s (s=1,...,$n_s$), while different cell lines are labeled by subindex α (α=1,...,$n_{cl}$). Since samples are described by sets of N miR expression values, they can be represented as dots in N-dimensional space (in our case, N=370). We renormalize the distribution of measured values for each miR (across the $n_s$ samples) to have zero mean and unit variance. For each sample, we measure: 1) the Euclidean distance (in 370-D miR measurement space after renormalization) between that sample and the centroid of the remaining replicates for the same cell line, $d_s^{in}$; 2) the Euclidean distance between that sample and the centroid of each of the other cell lines, $d_{s,\alpha}^{out}$. Notice that the centroid of a group of objects in N-dimensional space is the mean position obtained by averaging the coordinate vectors over all the objects in the group, akin to the definition of the center of mass in physics, if all the objects are of equal mass. Supporting Information Figure S1 shows a schematic representation of the definition of the distances $d_s^{in}$ and $d_{s,\alpha}^{out}$. The distance ratio (in logarithmic scale) between replicate r and cell line α is then $R_{s,\alpha}$= log $(d_{s,\alpha}^{out}/d_s^{in})$. Supporting Information Figures S2-S4 show distance ratios $R_{r,\alpha}/(R_{r,\alpha})_{max}$, which are normalized to have a maximum value of 1.

Supporting Information Figure S2 shows outlier analysis results for AML cell line data, where all normalized distance ratios between 24 cell line samples (vertical axis) and 7 different cell lines (horizontal axis) were calculated based on Euclidean distances in the 370-dimensional miR measurement space. The block-diagonal structure indicates that each sample lies closer to the centroid of the remaining replicates from the same cell line, relative to its distance to the centroid of each of the other cell lines. Therefore, we conclude that all AML replicates pass the robustness test. Supporting Information Figure S3 shows results for B-ALL cell line data (36 samples from 9 different cell lines), while Supporting Information Figure S4 displays results for T-ALL (20 samples from 5 different cell lines). In all cases, we observe similar block-diagonal matrix values that indicate that none of the measured samples should be dismissed as outlier.

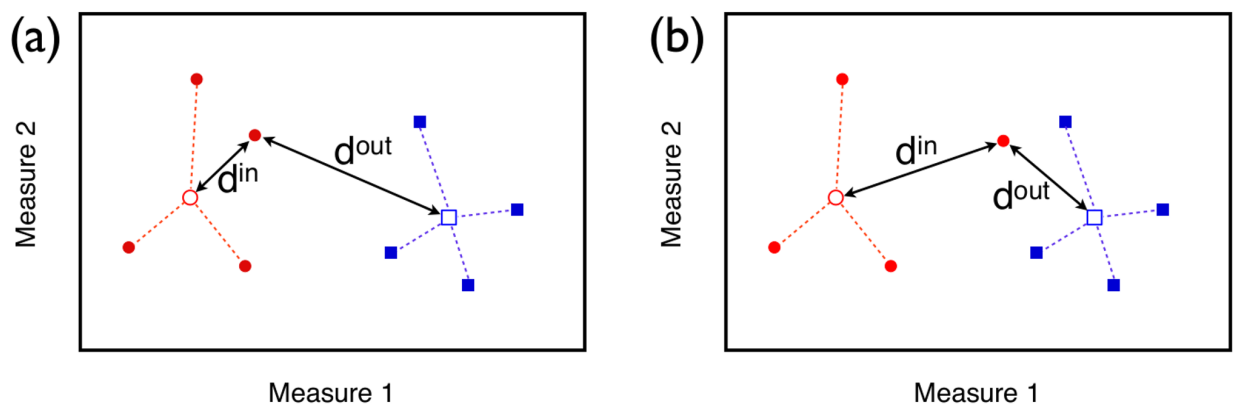

**CAPTION TO SUPPORTING INFORMATION FIGURE S1:**

**Schematic representation of the centroid-based method for outlier analysis.** Assuming for simplicity a 2-dimensional measurement space, each filled symbol represents a sample. Red circles represent replicates from cell line 1, blue squares represent replicates from cell line 2. The centroids for each group are shown by open symbols. The distance between the test sample and the centroid obtained from the remaining replicates of cell line 1 is labeled $d^{in}$. The distance between the test sample and the centroid obtained from all replicates of cell line 2 is labeled $d^{out}$. In **(a)**, $d^{in} < d^{out}$ and the chosen sample passes the robustness test. In **(b)**, $d^{in} > d^{out}$ and the chosen sample fails the test, therefore it is considered an outlier in the dataset.

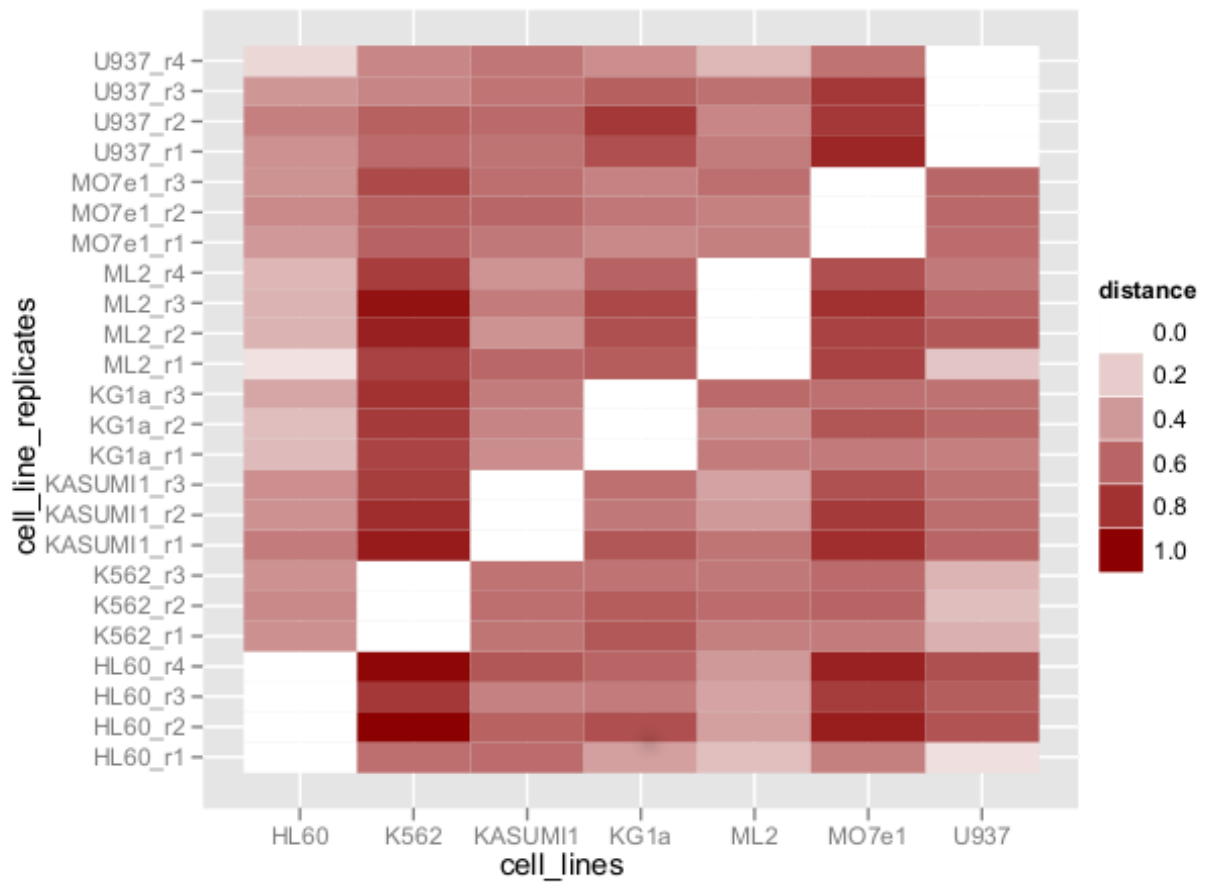

**CAPTION TO SUPPORTING INFORMATION FIGURE S2:**

**Outlier analysis of AML cell line data: 24 replicates from 7 cell lines.** Normalized distance ratios between cell line replicates (vertical axis) and cell lines (horizontal axis) based on Euclidean distances in the 370-dimensional miR measurement space. The block-diagonal structure indicates that each replicate lies closer to the center of mass of the remaining replicates from the same cell line, relative to its distance to the center of mass of each of the other cell lines.

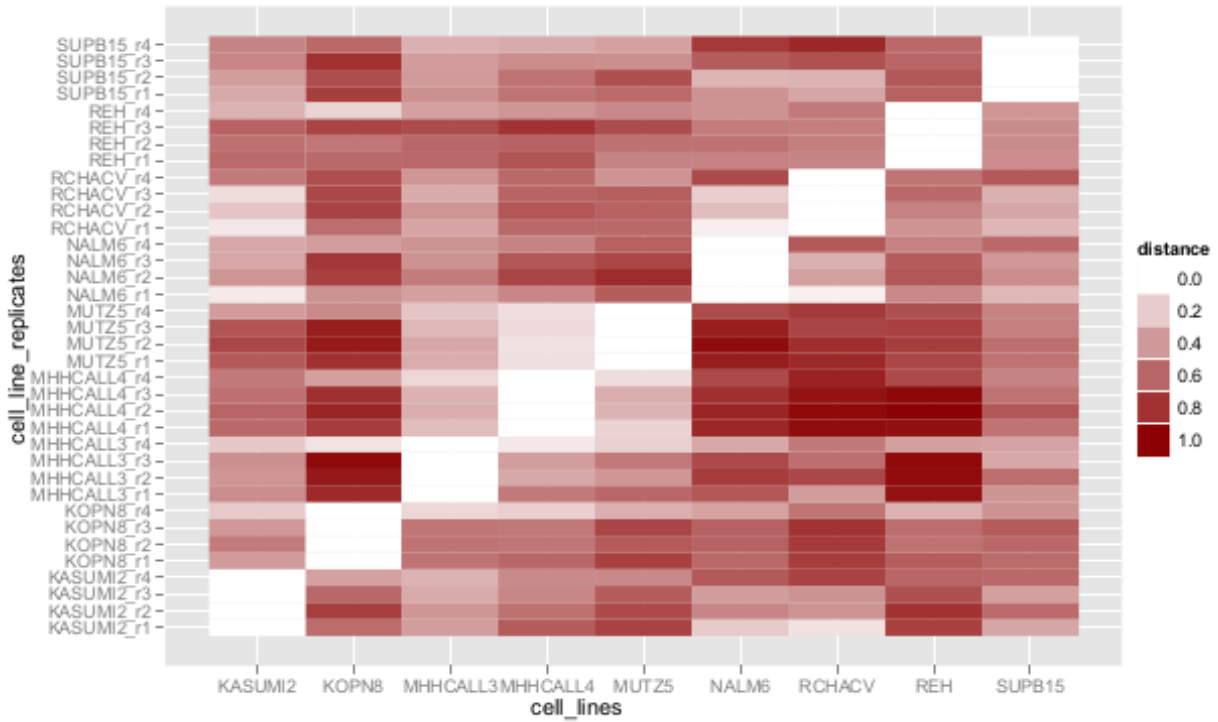

**CAPTION TO SUPPORTING INFORMATION FIGURE S3:**

**Outlier analysis of B-ALL cell line data: 36 replicates from 9 cell lines.** Normalized distance ratios between cell line replicates (vertical axis) and cell lines (horizontal axis) based on Euclidean distances in the 370-dimensional miR measurement space. As in Fig. S2, the block-diagonal structure indicates that all replicates pass the robustness test.

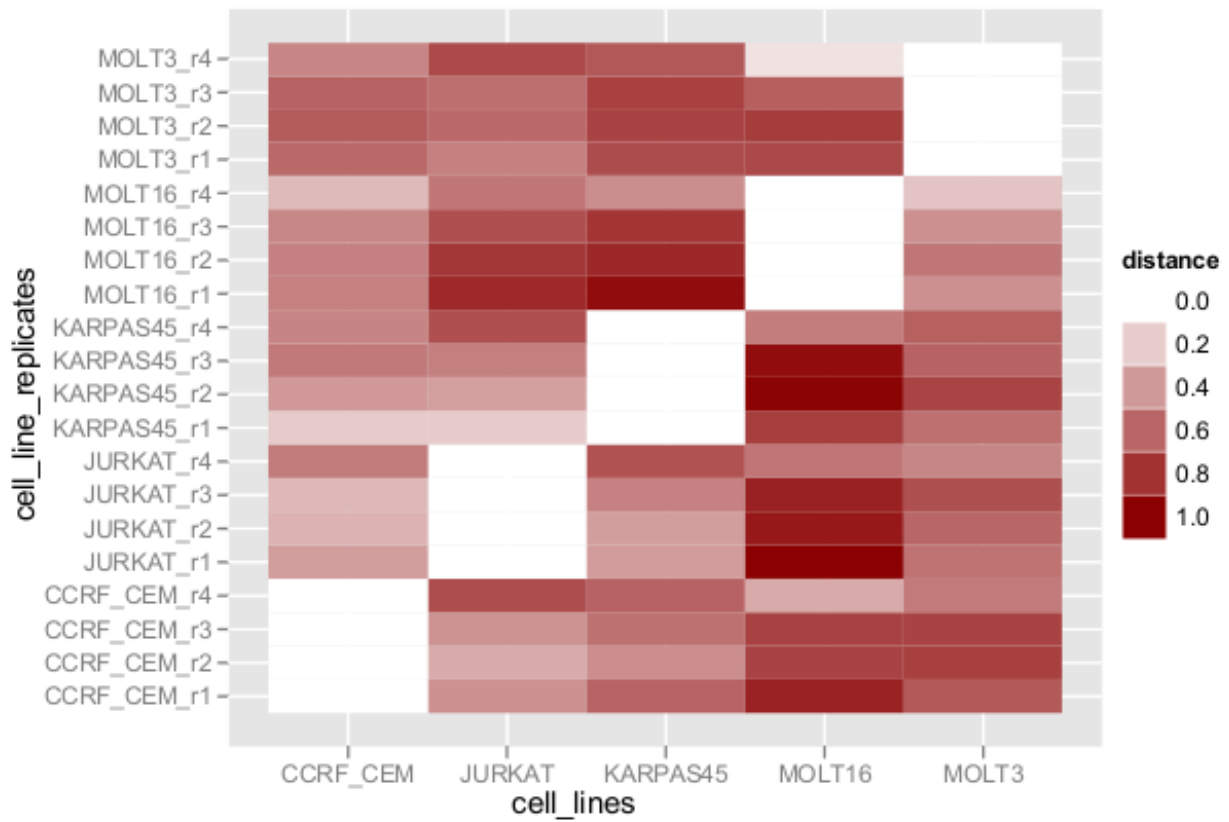

**CAPTION TO SUPPORTING INFORMATION FIGURE S4:**

**Outlier analysis of T-ALL cell line data: 20 replicates from 5 cell lines.** Normalized distance ratios between cell line replicates (vertical axis) and cell lines (horizontal axis) based on Euclidean distances in the 370-dimensional miR measurement space. As in Figs. S2 and S3, the block-diagonal structure indicates that all replicates pass the robustness test.